\documentclass[12pt]{article}
\usepackage{latexsym,epsfig}
\usepackage{graphicx}
\usepackage{pifont}
\usepackage{color}
\usepackage{verbatim}
\def\nn{\nonumber}
\def\ed{\end{document}}

\def\beq{\begin{eqnarray}}
\def\eq{\end{eqnarray}}
\def\beqn{\begin{eqnarray*}}
\def\eqn{\end{eqnarray*}}

\def\nl{\noindent}

\begin{document}
\begin{center}
{\bf \large Searching for an Extra Neutral Gauge Boson from Muon Pair Production at LHC } 
\vskip .5 cm
{E. Ramirez Barreto}

{Centro de Ci\^encias Naturais e Humanas, UFABC}

{Santo Andr\'e, SP, Brazil}
\vskip .5 cm

{Y. A. Coutinho }

{Instituto de F\'isica, UFRJ
,}

{Rio de Janeiro, RJ, Brazil}
\vskip .5 cm

{J. S\'a Borges}

{UERJ,}
{Rio de Janeiro, RJ, Brazil }
\end{center}

\begin{abstract}
We search  for signatures  of the extra neutral gauge boson $ Z^\prime$, predicted in some extensions of the Standard Model, from the analysis of some distributions for $p + p \longrightarrow \mu^+ + \mu^- + X$, where the only exotic particle involved is $ Z^\prime$. In addition to the invariant mass and charge asymmetry distributions,  we propose in our search  to use the transverse momentum distribution ($p_T$) as an observable.  We do our calculation for two values of the LHC  center of mass energy  ($7$ and $14$ TeV), corresponding to $1$ and $100$ fb$^{-1}$ of luminosity, in order to compare our findings from some models  with the distributions following from the Standard Model. By applying  convenient cuts in  the invariant mass, we  show that the final particles $p_T$ distributions can  reveal the presence of an extra neutral gauge boson contribution.  We also claim that  it is possible to disentangle the  models considered here  and we emphasize that the minimal version of the model, based on ${SU (3)_C \times SU (3)_L \times U (1)_X }$ symmetry, presents the more clear signatures for $ Z^\prime$ existence.

\vskip 0.6cm
\par
PACS: 12.60.Cn,14.70.Pw
\par

email: elmer@if.ufrj.br, yara@if.ufrj.br, saborges@uerj.br

\end{abstract}

\vfill\eject
\section{Introduction}

It is believed that the Standard Model (SM) of the electroweak
and strong interactions is not the complete theory because it does not explain some theoretical features, for example the family replication. This motivates the formulation of many  theoretical extensions of the SM. For example, the issue for the family replication problem and for the bound on the Weinberg angle is proposed in \cite{VIP}.

On the other hand the SM is in accordance with all available  experimental data from  LEP, SLD and Tevatron and, apart from the Higgs particle, all predicted particles have been  discovered.
It is  expected, in a near future,  that the Large Hadron Collider (LHC) at CERN will reach a higher energy and luminosity regime opening the possibility to find the responsible for electroweak symmetry breaking mechanism and to reveal phenomena related to new physics, such as the existence of new particles.

Many alternative models  predict the existence of new and exotic particles. Among them, the new neutral gauge  boson called $Z^\prime$ appears in the 3-3-1 models \cite{PIV, FRA, RHN}, little Higgs model \cite{LIT}, left-right symmetric models \cite{LRM}, superstring inspired E$6$ model \cite{E6P, E6M} and models with extra dimensions as Kaluza-Klein excitations of neutral gauge bosons \cite{RIZ}.

The search for $Z^\prime$ will play an important part in the proposals of future high-energy colliders. It is expected that LHC will be able to look for a $Z^\prime$ up to $5$ TeV, however only the next generation of linear colliders (ILC), dealing with polarized beams and  with $\sqrt s > 500$ GeV will be able to confirm its  existence {\it via} interference effects and to perform more precise measurements. 

 Unfortunately the $Z^\prime$  mass is not predicted in any model but the minimal version of 3-3-1 model  imposes an upper bound around $4$ TeV for this parameter what makes the model very  attractive under the experimental point of view. The acceptable current lower bound  is $600$ GeV and it is strongly model dependent \cite{PDG}. Previous results from interference  effects, at LEP, and from direct production, at the Tevatron, show that $Z^\prime$ is expected to be very heavy, having a small mixing with $Z$.  Some improvements on its mass and mixing have recently been obtained for some models from electroweak precision data  \cite{LAN}.

A safe way to search a $Z^\prime$ is to follow the same procedure adopted for discovering  $Z$ gauge boson. In this case, the searches must cover a range for an invariant dilepton mass higher than $M_Z$. The signature for its existence is obtained directly from proton-proton and $e^+ e^-$ collisions analyzing the distributions of $Z^\prime$ decay products. Once discovered the particle, one has to determine its main properties: mass, natural width, charge and spin. 

In order to confirm  the $Z^\prime$ existence, we have to collect a bulk of observable showing its signature. All these observable must come from the distributions of its decay products.  As emphasized by \cite{RIT}, the forward-backward  asymmetry (A$_{FB}$) final states distribution can reveal an axial coupling of $Z^\prime$ to fermions, then showing a clear signature. As it is 
known the $Z^\prime$ rapidity measurements can tell us about its couplings to quarks and leptons \cite{AGU}. 

In this paper we are proposing to consider another observable allowing for a $Z^\prime$ identification in proton-proton collisions at center of mass (c.o.m) energy  $\sqrt s = 14$ TeV with an annual luminosity of ${\cal {L}} = 100$ fb$^{-1}$. In addition, we present our results for a possible first stage of LHC operation, namely $\sqrt s = 7$ TeV  and  ${\cal {L}} = 1$ fb$^{-1}$. In this work we combine the forward-backward and invariant mass distributions with the transverse momentum distribution of the emerging fermions in  the processes $p + p \longrightarrow \mu^+ + \mu^- + X$ using some representative models containing $Z^{\prime}$. 

The present work is organized in the following way: in section 2 we show the couplings between $Z^\prime$ and quarks and leptons for some models, in the section 3 we give our results and finally  we present our conclusions.

\section {Models} 

\nl
In our study, we consider three different approaches to physics beyond the SM. One consider the GUT superstring-inspired $E_{6}$ and their low energy $\chi$ version, $SU(2)_{L}\times U(1)_{Y}\times U(1)_{Y^{\prime}}$. Another approach implies an extension of the SM like the left-right models that appear in order to explain the left-right parity  breaking and is based in the gauge group $SU(2)_{L}\times SU(2)_{R} \times U(1)_{B-L}$. Finally, we consider an alternative approach to the SM, represented by the models with symmetry $SU(3)_{C}\times SU(3)_{L} \times U(1)_{X}$, called 3-3-1 models for short, that give answer to some open questions as the number of families or the electroweak mixing angle value. The fermion-$Z^{\prime}$ couplings are specific for each model. This characteristic is reflected in the width of the new neutral gauge boson, that define a leptophobic or leptophilic character of $Z^{\prime}$.

The general Lagrangian for the neutral current involving $Z$ and $Z^{\prime}$ contributions for the models studied in this article is: 
\bigskip
\beq
&&{\cal L}^{NC} =-\frac{g}{2 \cos\theta_W}\sum_{f} \Bigl[\bar
f\, \gamma^\mu\ (g_V + g_A \gamma^5)f \, Z_\mu+ \bar f \, \gamma^\mu\ (g^\prime_V + g^\prime_A \gamma^5)f \, { Z_\mu^\prime}\Bigr],
\nn
\eq
where $f$ can be leptons and quarks and $g$ is the weak coupling constant,
$g_V$ and $g_A$ are the SM  couplings, whereas the new couplings  $g^\prime_V$ and  $g^\prime_A$ are presented in the Tables I and II.
Below the electroweak scale, the phenomenology predicted by these models involving the $\gamma$ and the Z, coincides with  the SM one.

\begin{table}[h]\label{sagui}
\begin{footnotesize}
\begin{center}
\begin{tabular}{||c|c|c|c|c||}
\hline \hline
\multicolumn{3}{|c|}{3-3-1 MIN} &
\multicolumn{2}{|c|}{3-3-1 RHN}
\\ 
\hline
&  $g^\prime_V$ & $g^\prime_A$   & $g^\prime_V$ & $g^\prime_A$ \\
 \hline
$Z^{\prime} \bar l l $ &
$\displaystyle{-\frac{\sqrt 3}{2}{\sqrt{1-4\sin^2\theta_W}}}$ &
$\displaystyle{\frac{\sqrt 3}{6}{\sqrt{1-4\sin^2\theta_W}}}$ &
$\displaystyle{\frac{-1+4\sin^2\theta_W}{2\sqrt{3-4\sin^2\theta_W}}}$ &
$\displaystyle{-\frac{1}{2\sqrt{3-4\sin^2\theta_W}}}$ \\
\hline
$Z^{\prime} \bar u u$ & $\displaystyle{-\frac{1+4\sin^2\theta_W}{2\sqrt{3-12\sin^2\theta_W}}}$
& $\displaystyle{\frac{1}{\sqrt{3-12\sin^2\theta_W}}}$ &  $\displaystyle{\frac{3-8\sin^2\theta_W}{{6\sqrt{3-4\sin^2\theta_W}}}}$  & $\displaystyle{-\frac{1}{2\sqrt{3-4\sin^2\theta_W}}}$  \\
\hline
$Z^{\prime} \bar d d$ &  
$\displaystyle{\frac{1-2\sin^2\theta_W}{2\sqrt{3-12\sin^2\theta_W}}}$  &  
$\displaystyle{-\frac{1+2\sin^2\theta_W}{2\sqrt{3-12\sin^2\theta_W}}}$  &  
$\displaystyle{\frac{3- 2\sin^2\theta_W}{6\sqrt{3-4\sin^2\theta_W}}}$   &  $\displaystyle{-\frac{\sqrt{3-4\sin^2\theta_W}}{6}}$  \\
\hline
\end{tabular}
\end{center}
\end{footnotesize}
\caption{The vector and axial couplings of $Z^{\prime}$ with leptons ($e$, $\mu$ and $\tau$) and quarks ($u$ and $d$) in the 3-3-1 models.  $\theta_W$ is the Weinberg angle.}
\end{table}

\begin{table}[h]\label{mico}
\begin{footnotesize}
\begin{center}
\begin{tabular}{||c|c|c|c|c||}
\hline \hline
\multicolumn{3}{|c|}{Sym L-R} &
\multicolumn{2}{|c|}{$E_{6}-\chi$}
\\ 
\hline
&  $g^\prime_V$ & $g^\prime_A$   & $g^{\prime}_V$ & $g^{\prime}_A$ \\
\hline
$Z^{\prime} \bar l l $ &
$\displaystyle{\frac{-1+4\sin^2\theta_W}{2 \sqrt{\cos^2\theta_W - \sin^2\theta_W}}}$ &
$\displaystyle{\frac{\sqrt{\cos^2\theta_W-\sin^2\theta_W}}{2}}$ &
$\displaystyle{\frac{2 \sin\theta_W}{\sqrt 6}}$ &
$\displaystyle{\frac{\sin\theta_W}{\sqrt 6}}$ \\
\hline
$Z^{\prime} \bar u  u$ & $\displaystyle{\frac{3-8\sin^2\theta_W}{6\sqrt{\cos^2\theta_W - \sin^2\theta_W}}}$
& 
$\displaystyle{-\frac{\sqrt{\cos^2\theta_W-\sin^2\theta_W}}{2}}$ 
&  $\displaystyle{0}$  & $\displaystyle{\frac{\sin\theta_W}{\sqrt 6}}$  \\
\hline
$Z^{\prime} \bar d d$ &  
$\displaystyle{\frac{-3+4\sin^2\theta_W}{6\sqrt{\cos^2\theta_W - \sin^2\theta_W}}}$  &
$\displaystyle{\frac{\sqrt{\cos^2\theta_W-\sin^2\theta_W}}{2}}$  &
  $\displaystyle{-\frac{2 \sin \theta_W}{\sqrt 6}}$   &  $\displaystyle{-\frac{\sin\theta_W}{\sqrt 6}}$  \\
\hline
\end{tabular}
\end{center}
\end{footnotesize}
\caption{The vector and axial couplings of $Z^{\prime}$ with leptons ($e$, $\mu$ and $\tau$) and quarks ($u$ and $d$) in the  Sym L-R  and $E_{6}-\chi$ models.}
\end{table}

\section{Results}

We calculate the invariant mass, forward-backward asymmetry and transverse momentum distributions for $p + p \longrightarrow \mu^+ + \mu^- + X$, at $14$ TeV, from the following models: a model with symmetry ${SU (3)_C \times SU (3)_L \times U (1)_X }$, in its  minimal version ($331-$MIN) \cite{PIV, FRA} and the version with right-handed neutrinos ($331-$RHN) \cite{RHN}, a left-right symmetric model (LRM) \cite{LRM} and $E_6$ inspired  model (E$_6 \, \chi$) \cite{E6P, E6M}. By respecting the different group assignments, it is clear that in  each model there is a peculiar  $Z^\prime$  coupling to fermions. All calculations of this work are performed with the CompHep package \cite{HEP}.

Considering  the Tevatron results and from the theoretical constraints on the considered models, we vary the $Z^\prime$ mass from  $800$ GeV to $1200$ GeV and we compare the distributions with the SM ones 

As a  first step to search its signature we study the invariant mass distributions. In this case, the determination  of $Z^\prime$ mass can become more easy if one consider convenient cuts in the invariant mass: ( $M_{\mu\, \mu} > 500$ GeV). In accordance with the CMS detector performance simulation \cite{CMS}, we adopted the following dimuon pseudo-rapidity and transverse momentum cuts: $\vert \eta\vert < 2.5$ and  $p_T > 20$ GeV. 
In the Figures 1 we display the resulting muon pair mass distributions. One can observe that the shape of the distributions are related to the $Z^\prime$ width. For the $331-$MIN, there are many $Z^\prime$ decay channels available and so the distribution is more flat than the others. A remarkable observation is that, even for small dimuon mass values, the  predicted number of dimons in $331-$MIN, are much larger than those predicted by the other models, including the SM one.

Next we study the angular distributions, in order to analyze the forward-backward asymmetry (A$_{FB}$). As pointed out in \cite{GOD} and \cite{ROS, DIT, DIM} this analysis is essential to disentangle $Z^\prime$ predicting models. It is well known that this calculation is quite difficult because the direction of the interacting quarks are not determined. To face this difficulty one uses the kinematics of the dimuon system. As the elementary processes involve quarks with various momentum distributions, one can  approximate the quark direction by the boost direction connecting the dimuon system with the beam axis. As a consequence the assignment of quark direction can be obtained if one selects dimuon large rapidity events. This way, we adopt the same muon rapidity cut ($\vert y_{\mu \mu} \vert > 0.8$) as proposed in \cite{DIT, DIM}. 
We present in the Figures 2 our results for the A$_{FB}$, by considering the dimuons invariant mass around the values of $800$, $1000$ and $1200$ GeV. These Figures show that the asymmetry calculated from all models, except the $331-$MIN model, is very sensitive to the  $Z^\prime$ mass. A$_{FB}$ from the $331-$MIN model is almost constant and this behavior is related to the leptophobic character of the $331-$MIN $Z^\prime$ particle.  Clearly the A$_{FB}$ analysis can be used to disentangle the  models.  The A$_{FB}$ analysis for some models from reconstruction of fully-simulated events was performed in \cite{CMS1}. 

In order to emphasize the role played by the exchange of an exotic neutral gauge boson, we have also calculated the fermion transverse momentum distribution obtained by integrating the elementary differential cross section over the fermion transverse momentum. The elementary cross section presents propagator poles  at the resonance masses ($Z$ and $Z^\prime$). In the integral, the Jacobian of the transformation, from the fermion momentum to its transverse component, introduces the factor \cite{VER, JOH}
$$\sqrt{\frac{\hat s }{\hat s - 4 p_T^2}},$$
on the other hand, the more important contribution for the final state distribution comes when the energy of the elementary process is close to the resonance poles. As a consequence, the final momentum transverse distribution is more pronounced for $p_T = M_{res}/2$. As a consequence the transverse momentum distribution can be used to identify the resonance masses.

This way, two peaks are expected to appear in the muon $p_T $ distribution  corresponding to one half of the resonance masses ($M_Z$ and $M_{Z^\prime}$).
In order to have a more clear signature of $Z^\prime$ exchange we have applied a more strong cut in the dimuon mass ($M_{\mu\, \mu} > 500$ GeV) and we present in the Figures 3 our findings for $M_{Z^\prime}= 800, 1000$ and $1200$ GeV, respectively.  We observe in this case that, for all models including the SM one, there is a peak at $250$ GeV, which is related to the adopted invariant mass cut. The second peak keeps its position at $p_T = M_{Z^\prime}/2$. Moreover, we observe a complete different shape for the distributions: the LRM exhibits a more pronounced peak while the $331-$MIN is flat due to its peculiar $Z^\prime$ width.  We claim that this behavior can be used as a nice signature for $Z^\prime$ existence as well as it can be used to disentangle the different models.

We extended all calculations for the c.o.m. energy of $7$ TeV that corresponds to the initial operation condition of the LHC \cite{JEL}, and  the corresponding invariant mass and transverse moment distributions are displayed in the Figures 4 and 5. We observe that the distribution behavior are similar to that obtained for  $\sqrt 14$ TeV with a smaller expected number of events.

In conclusion we show that, in addition to $A_{FB}$ asymmetry,  the muon $p_T$ distribution from muon pair production is a very promising tool for LHC experimental groups (CMS and ATLAS) to discover the seed for new physics,  and for theoreticians,  to disentangle models. It is important to mention that the annual number of events in a possible first stage of LHC operation ($\sqrt s=7$ TeV and  
${\cal L} = 1$ fb$^{-1}$) are expected to be in the range 40 to 140 for $M_{Z^\prime}= 1200$ GeV  for all models studied in this work. On the other hand the number of events are predicted to be in the range from $3\times 10^4$ to $12\times 10^4$ for $\sqrt 14$ TeV and ${\cal L} = 100$ fb$^{-1}$ for the range of mass studied in our work.

Finally, we emphasize that, in all calculations,  we have adopted a realistic approach by considering, for each model, the "correct" $ Z^\prime$ width,  as  shown in the Table 3.

\vskip 1cm
\textit{Acknowledgments:} 
E. Ramirez Barreto thanks Capes. J. S\'a Borges and Y. A. Coutinho thank FAPERJ for financial support.

\begin{table}\label{manga}
\begin{center}
\begin{tabular}{||c|c|c|c|c||}
\hline
$M_{Z^\prime}$  & $\Gamma_{Z^\prime}$ ($331$-MIN)& $\Gamma_{Z^\prime}$ ($331$-RHN)  & $\Gamma_{Z^\prime}$ (LRM)  & $\Gamma_{Z^\prime}$ ($E_6$ $\chi$) \\ \hline
800 & $110$  &  $18$  & $17$  & $9$ \\ 
1000 & $186$  &  $23$  & $21$  & $12$ \\ 
1200 & $292$  &  $27$  & $25$  & $14$ \\
\hline
\end{tabular}
\end{center}
\caption{Some $M_{Z^\prime} $ values and their respectives widths in GeV for some models.}
\end{table}

\begin{figure}
\includegraphics[height=.3\textheight]{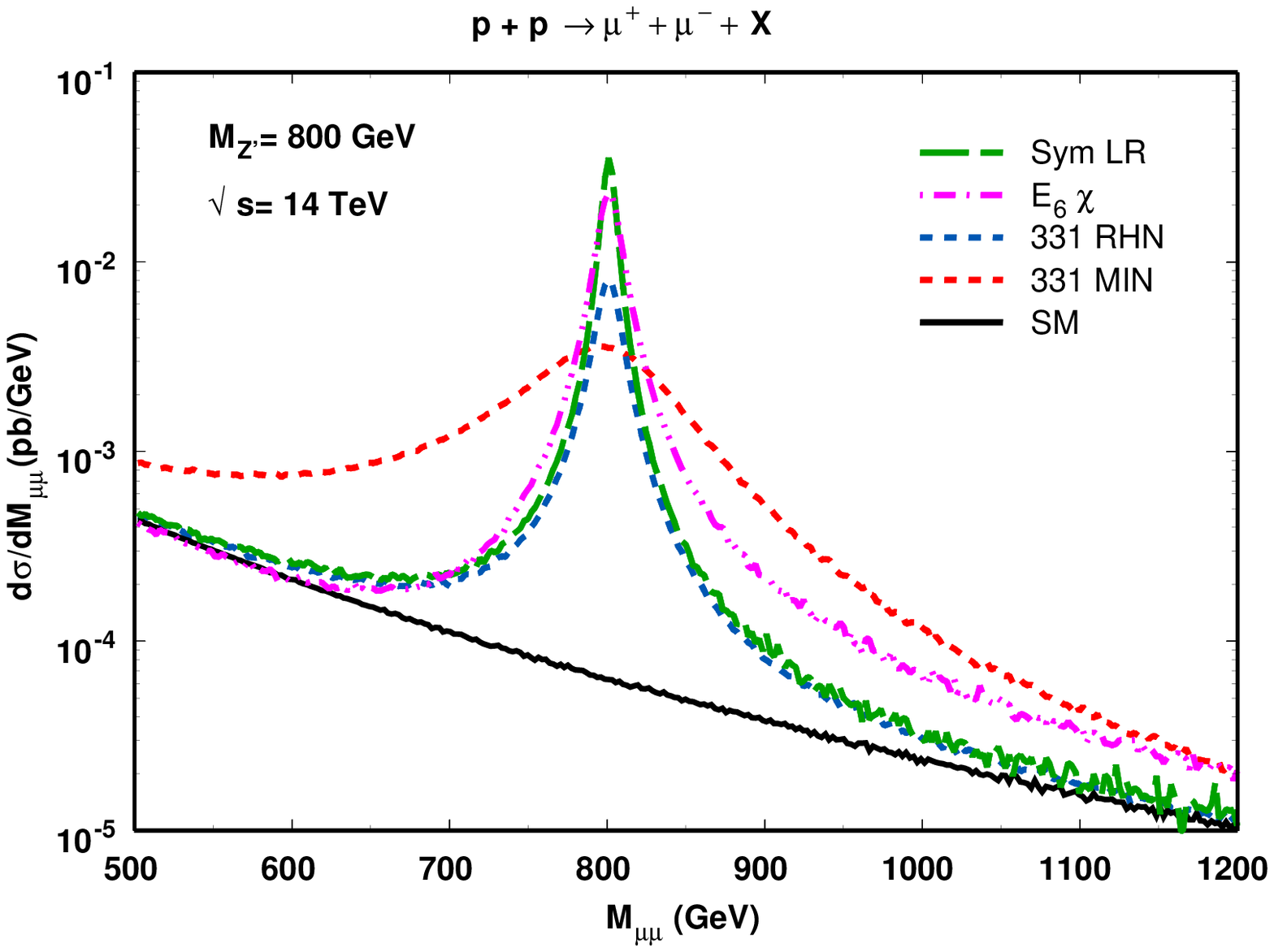}
\hskip -1. cm \includegraphics[height=.3\textheight] {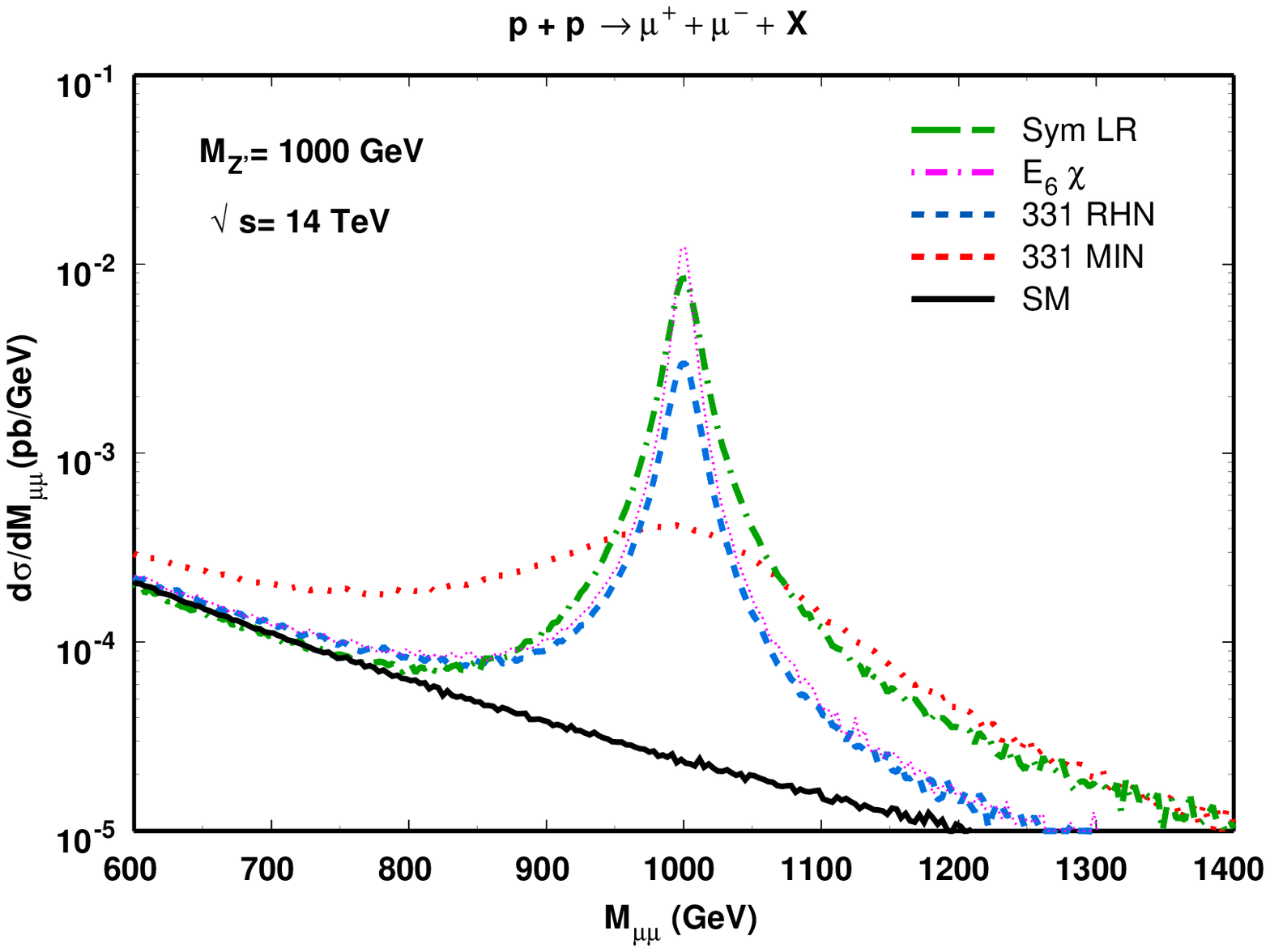}
\hskip -1. cm \includegraphics[height=.3\textheight] {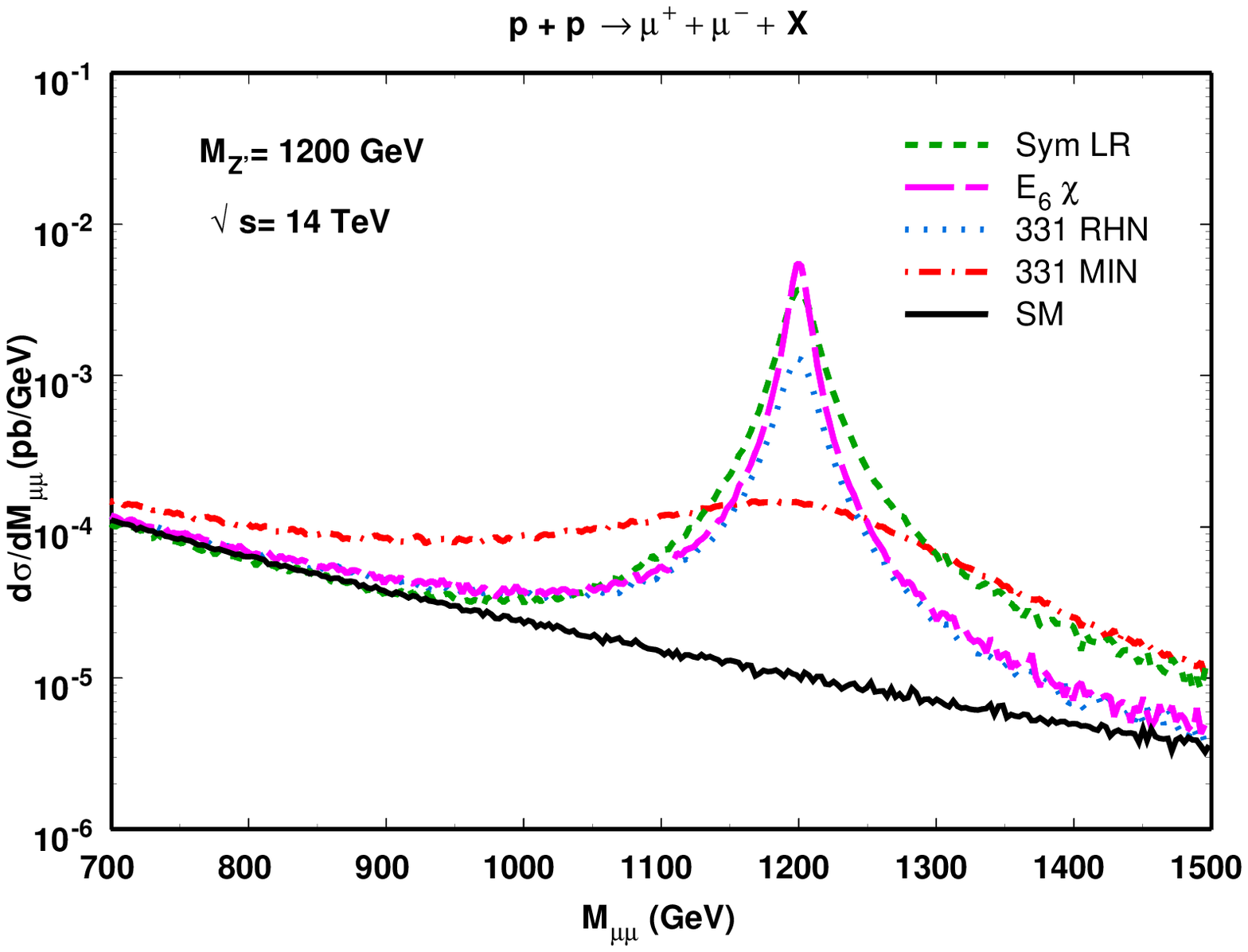}
\caption{\large Invariant mass distribution for the process $p + p \longrightarrow \mu^+ + \mu^- + X$ ($\sqrt s = 14$ TeV) for some models considering $M_{Z^\prime}=800 $ GeV (left), 
$M_{Z^\prime}=1000 $ GeV (right) and $M_{Z^\prime}=1200 $ GeV (down).}
\end{figure}

\begin{figure}
\includegraphics[height=.3\textheight]{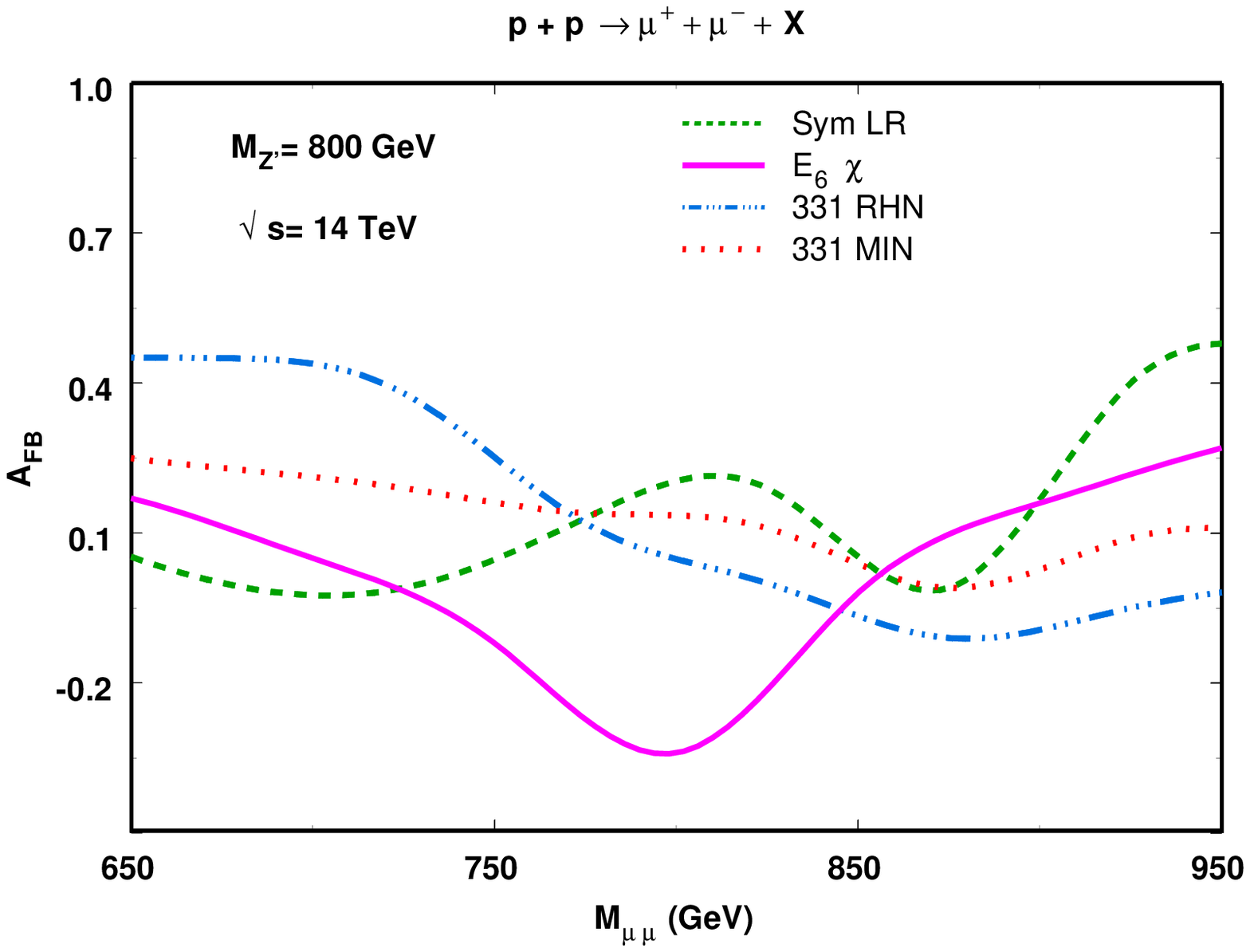}
\hskip -1. cm \includegraphics[height=.3\textheight] {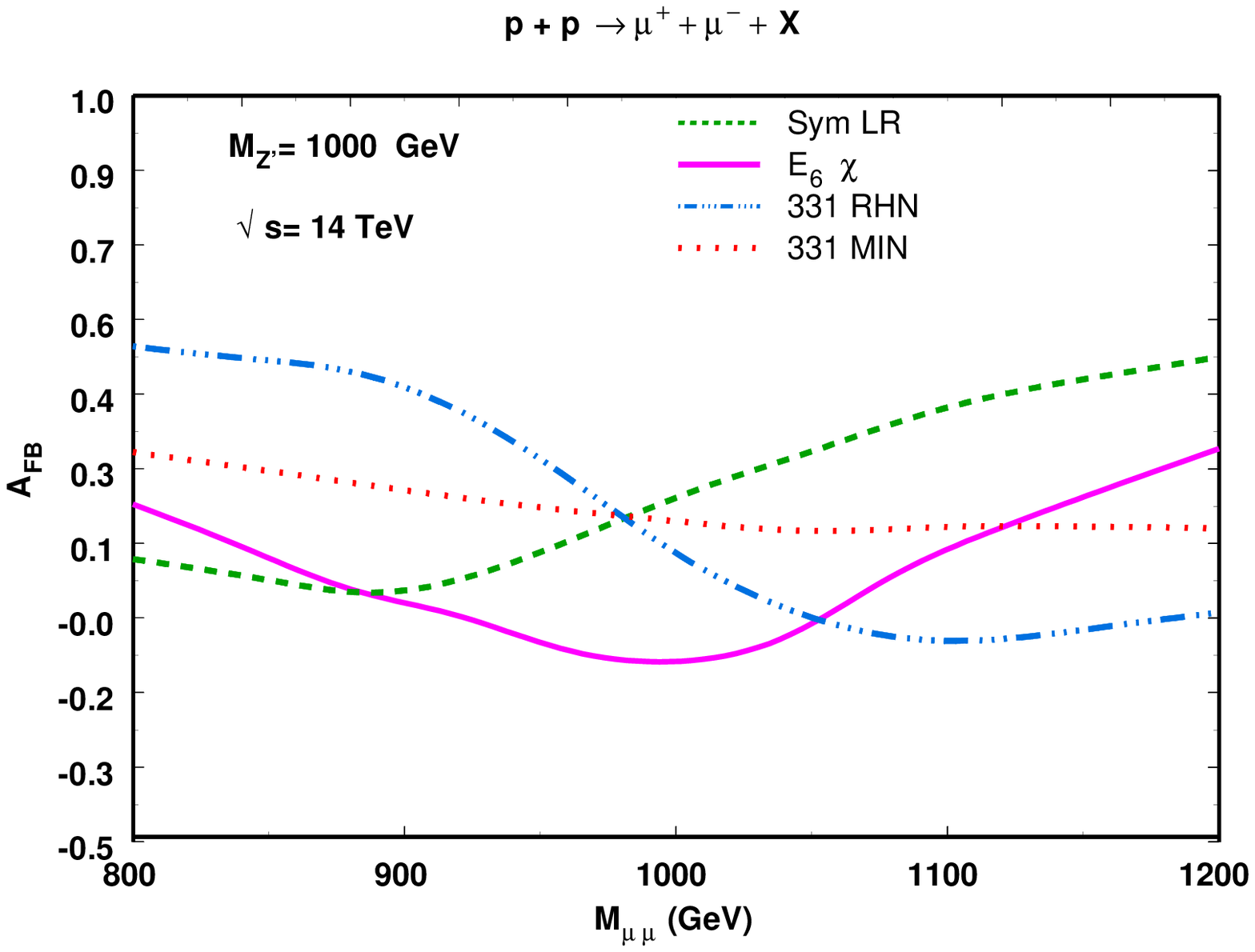}
\hskip -1. cm \includegraphics[height=.3\textheight] {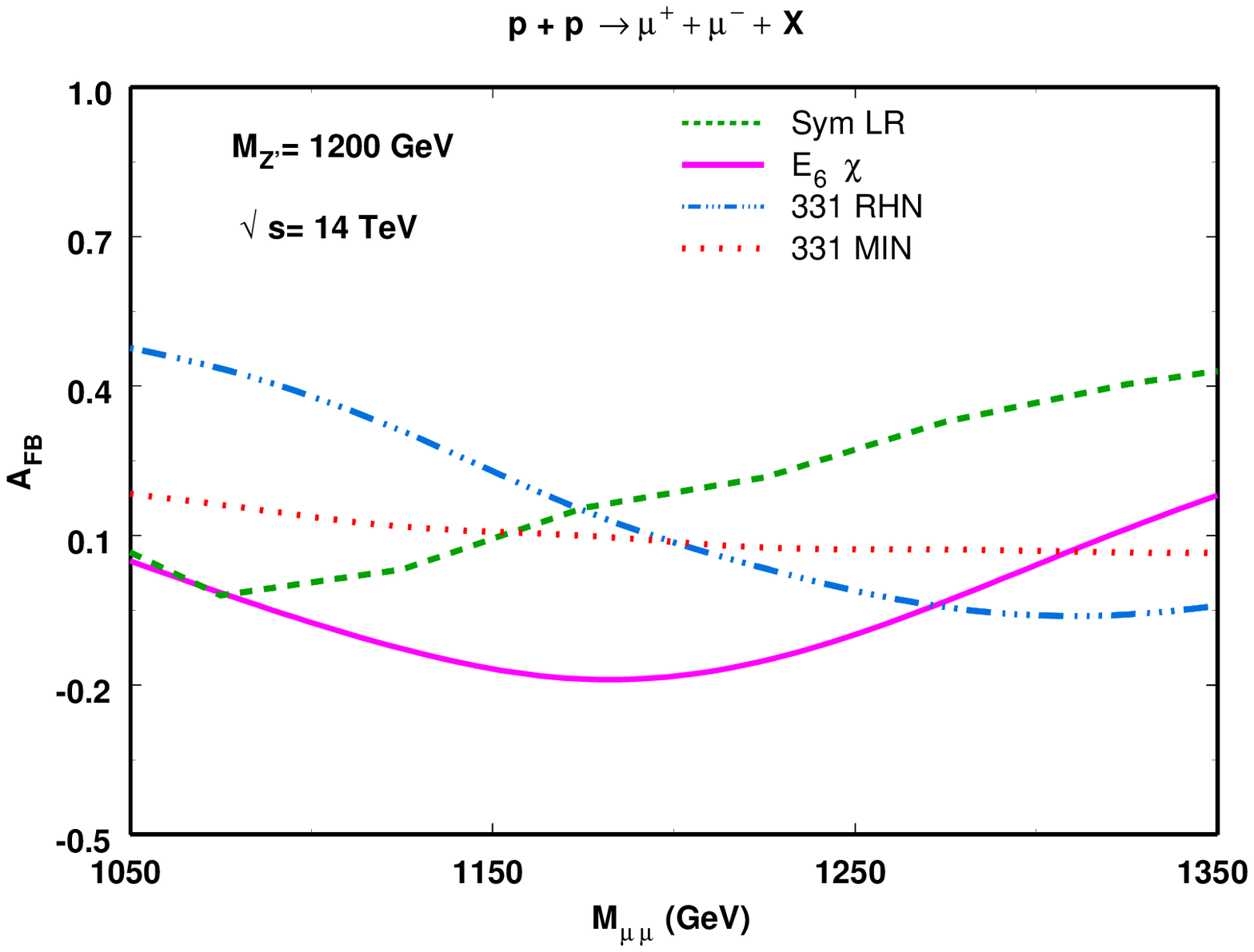}
\caption{\large Forward-backward asymmetry for the process $p + p \longrightarrow \mu^+ + \mu^- + X$ ($\sqrt s = 14$ TeV) for some models considering $M_{Z^\prime}=800 $ GeV (left), 
$M_{Z^\prime}=1000 $ GeV (right) and $M_{Z^\prime}=1200 $ GeV (down).}
\end{figure}

\begin{figure}
\includegraphics[height=.3\textheight]{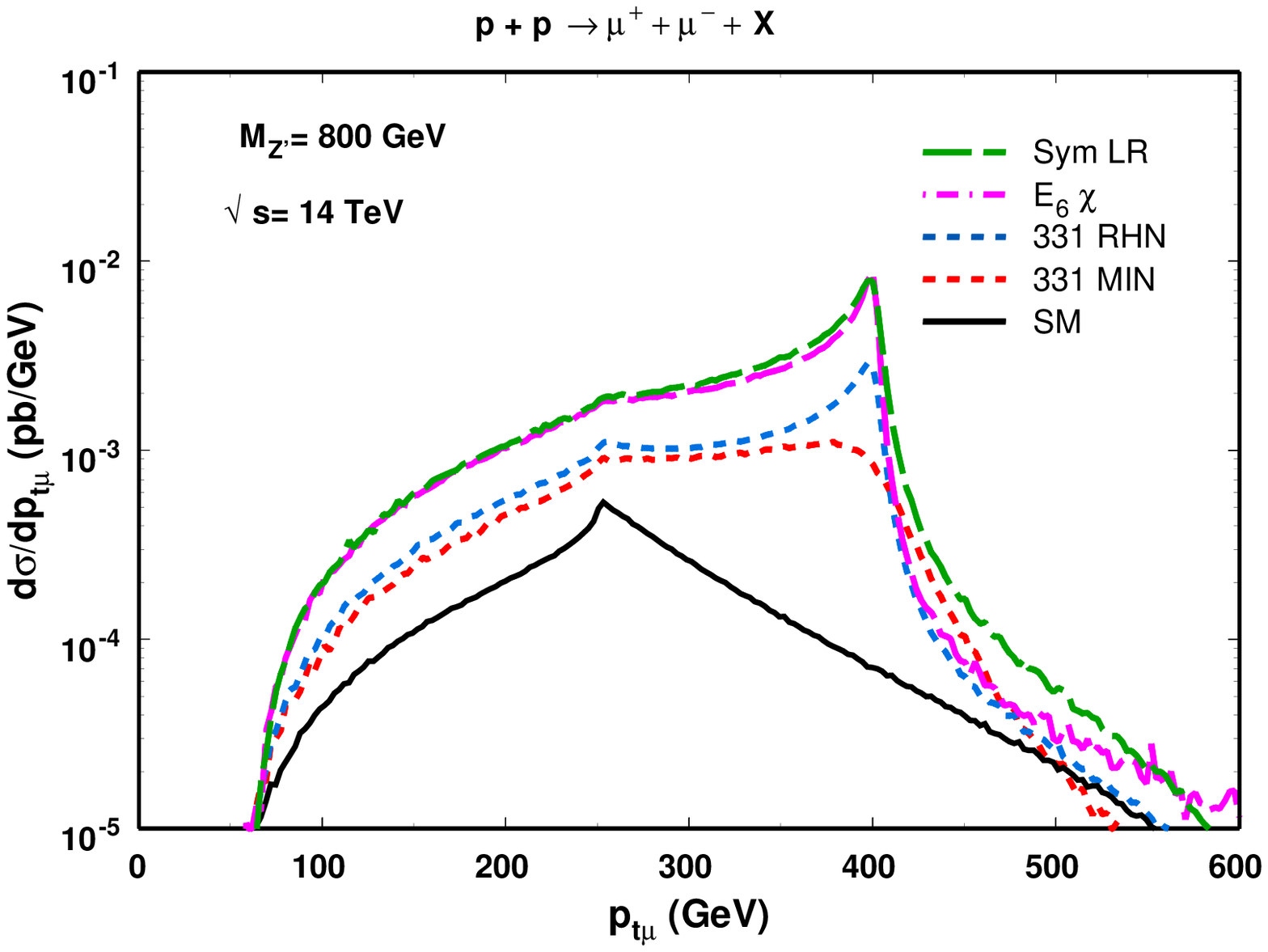}
\hskip -1. cm \includegraphics[height=.3\textheight] {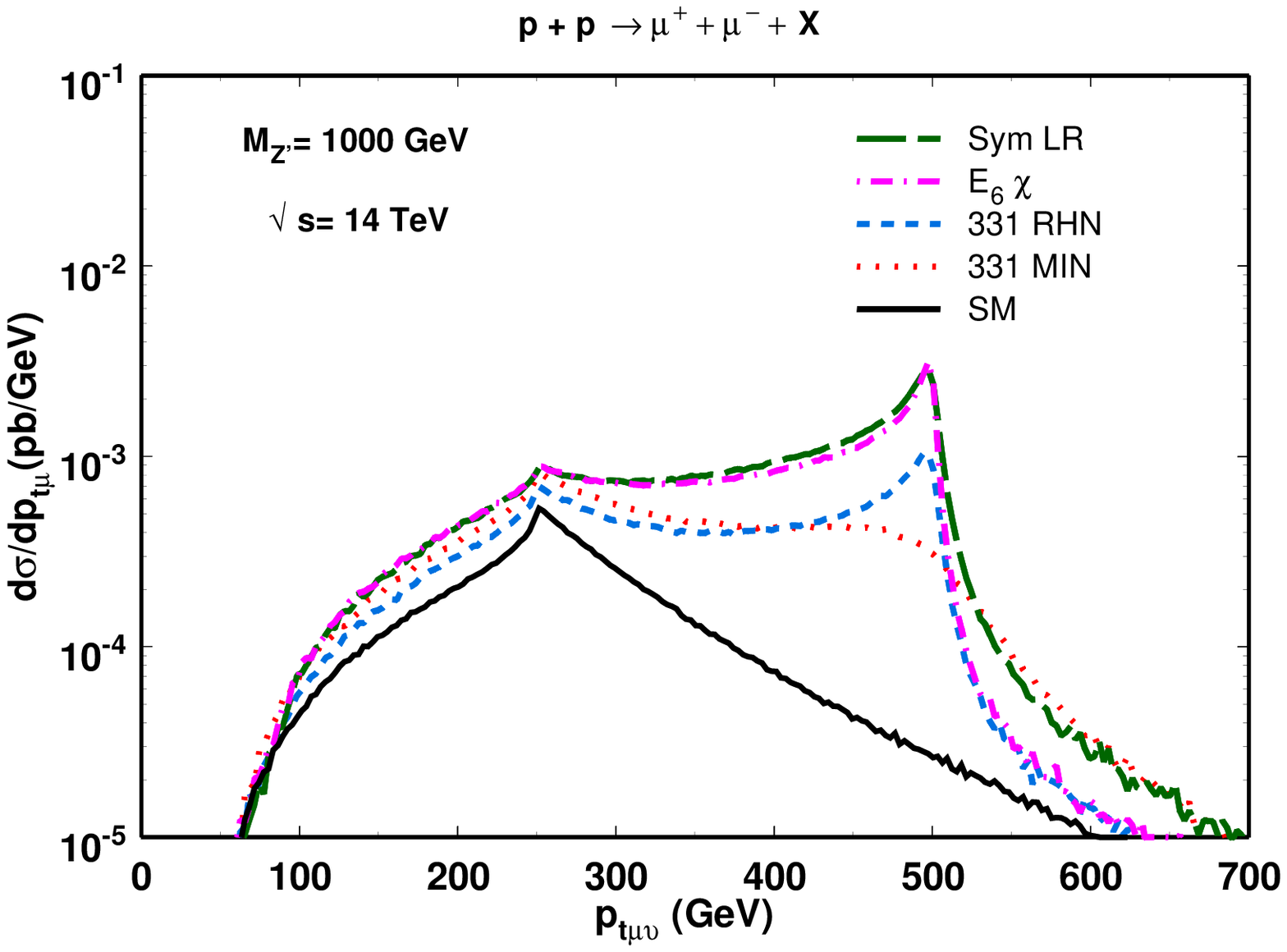}
\hskip -1. cm \includegraphics[height=.3\textheight] {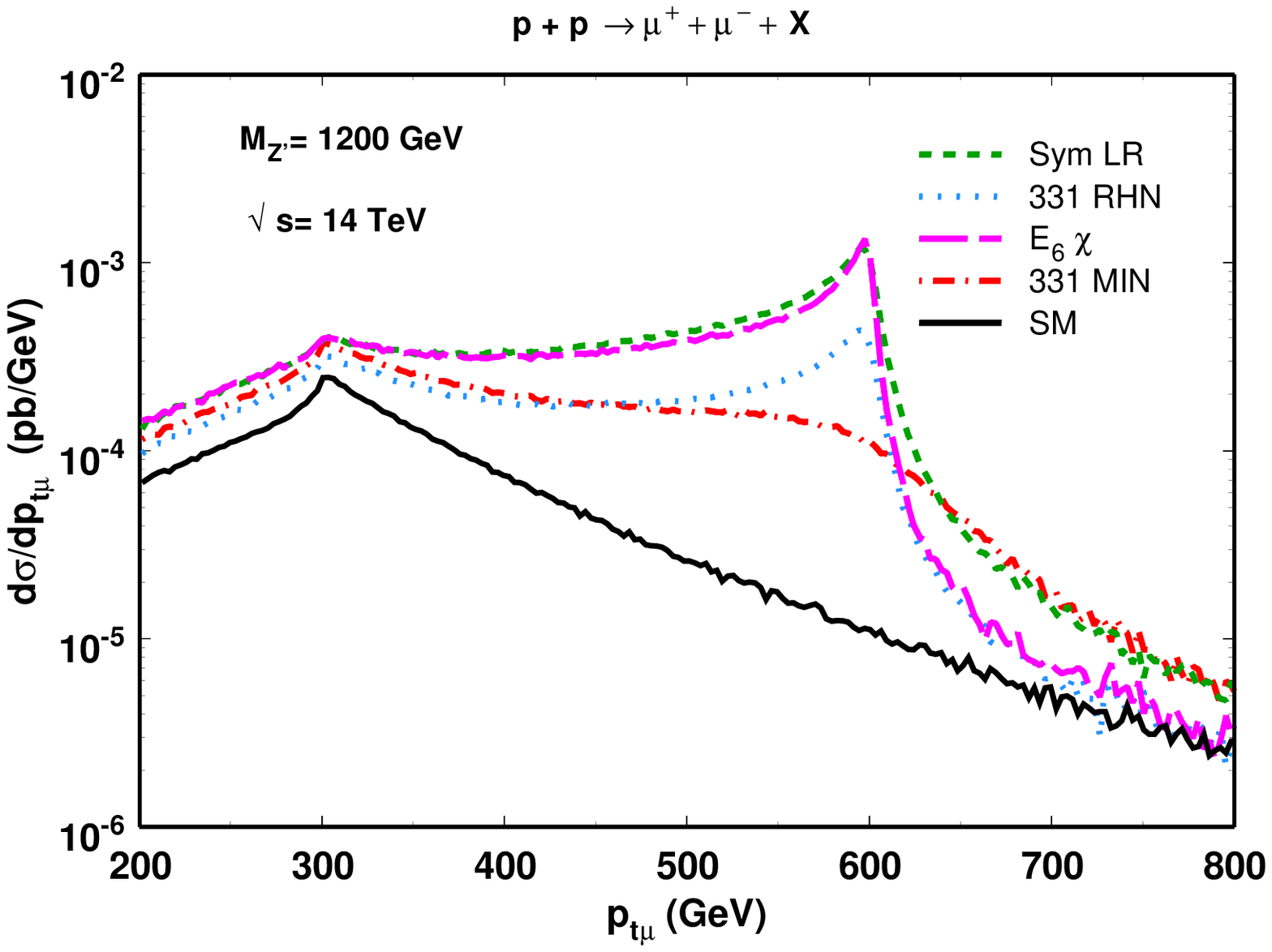}
\caption{\large Lepton transverse momentum distribution for the process $p + p \longrightarrow \mu^+ + \mu^- + X$ ($\sqrt s = 14$ TeV) for some models, considering $M_{Z^{\prime}}=800$ GeV (left), $M_{Z^\prime}=1000 $ GeV (right) and $M_{Z^\prime}=1200 $ GeV (down) and a cut $M_{\mu \, \mu} > 500 $ GeV.}
\end{figure}

\begin{figure}
\includegraphics[height=.3\textheight]{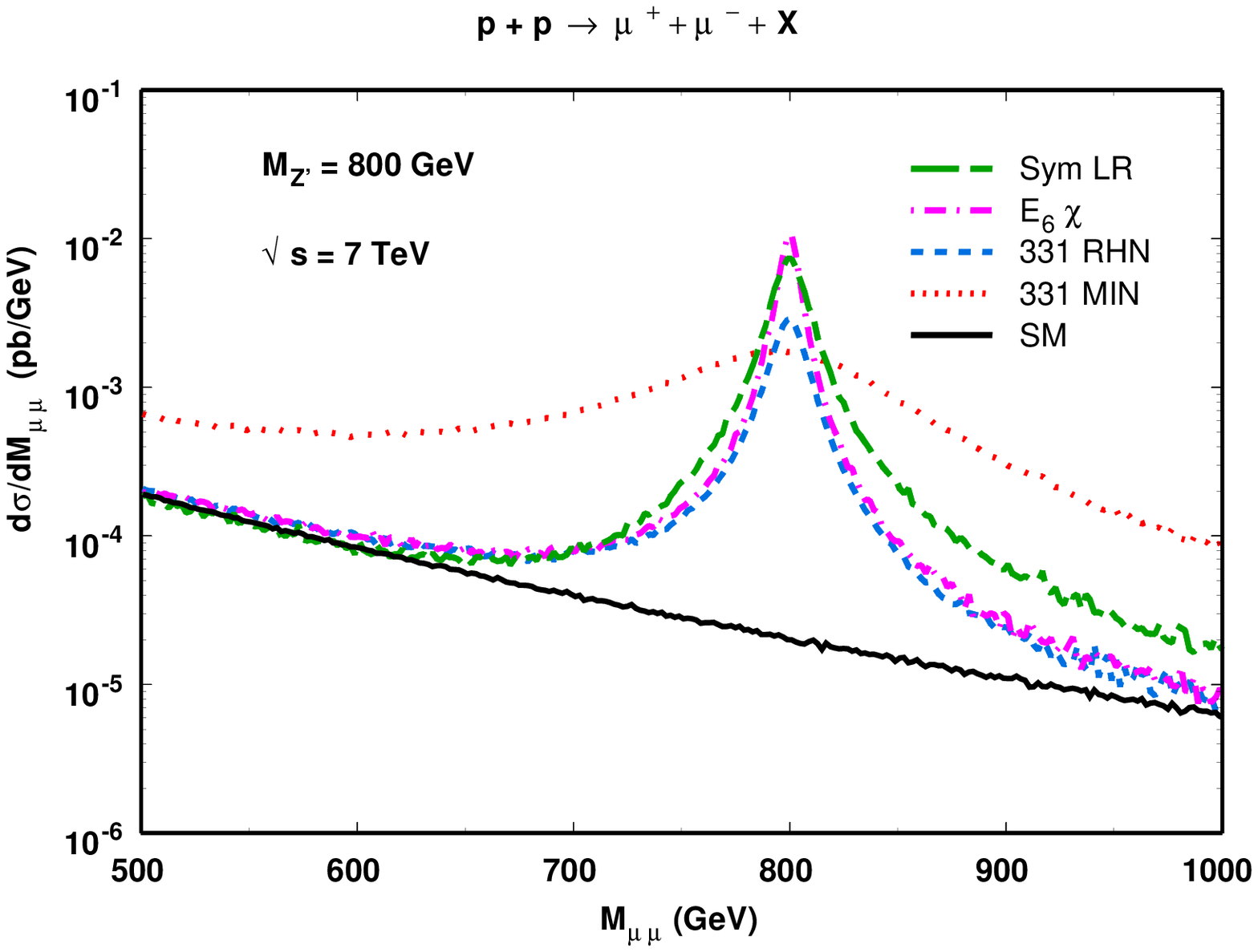}
\hskip -1. cm \includegraphics[height=.3\textheight] {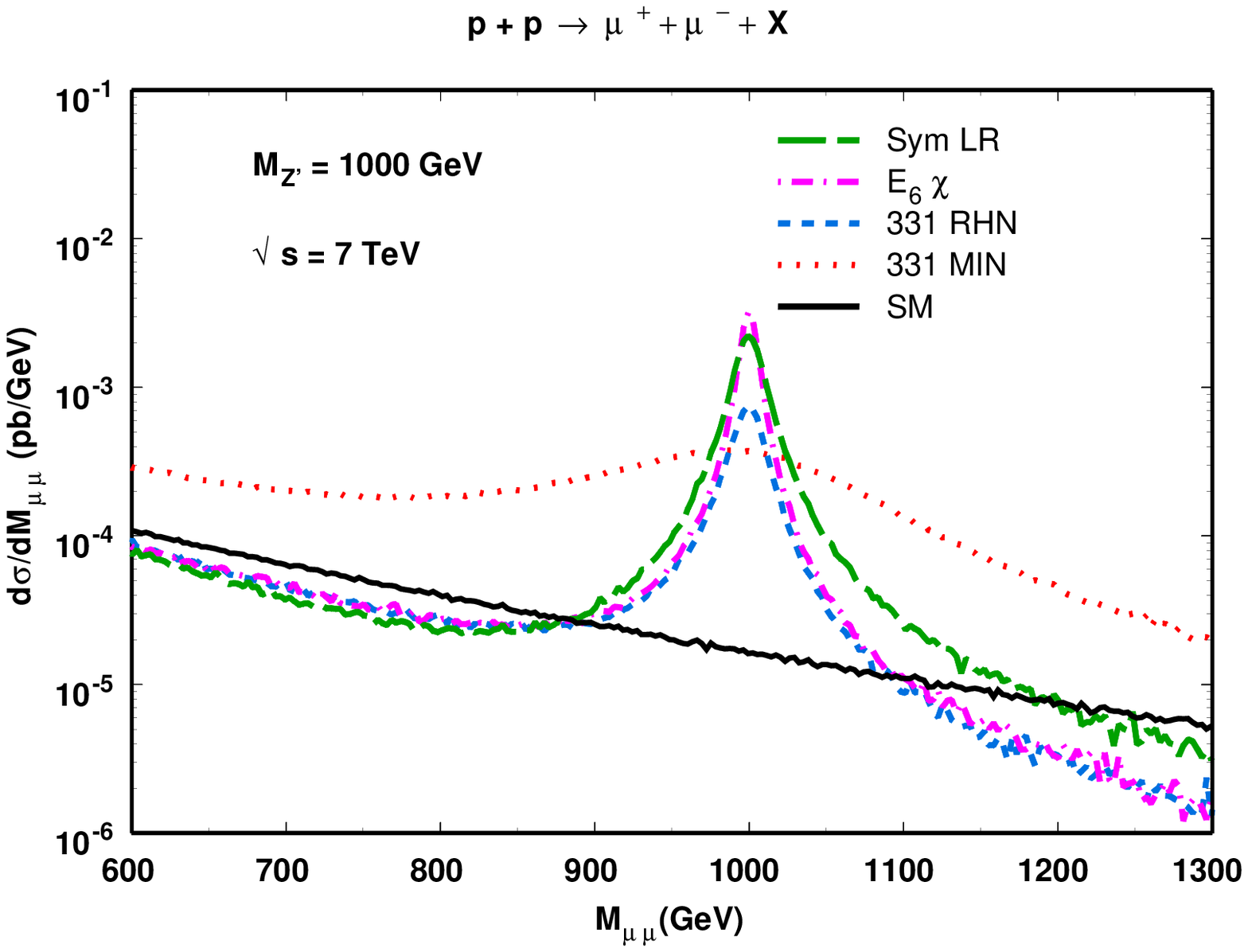}
\hskip -1. cm \includegraphics[height=.3\textheight] {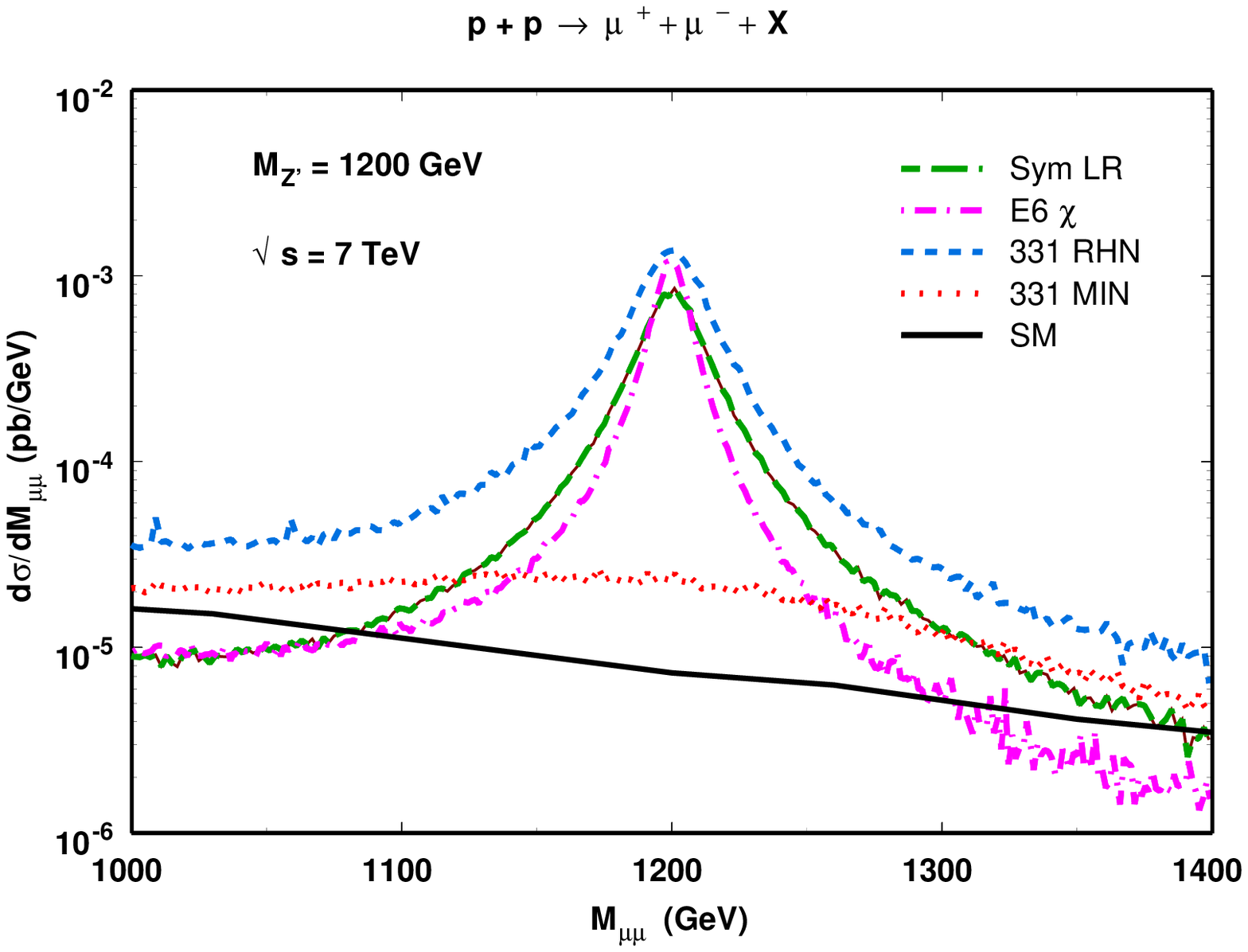}
\caption{\large Invariant mass distribution for the process $p + p \longrightarrow \mu^+ + \mu^- + X$ ($\sqrt s = 7$ TeV) for some models considering $M_{Z^\prime}=800 $ GeV (left), 
$M_{Z^\prime}=1000 $ GeV (right) and $M_{Z^\prime}=1200 $ GeV (down).}
\end{figure}

\begin{figure}
\includegraphics[height=.3\textheight]{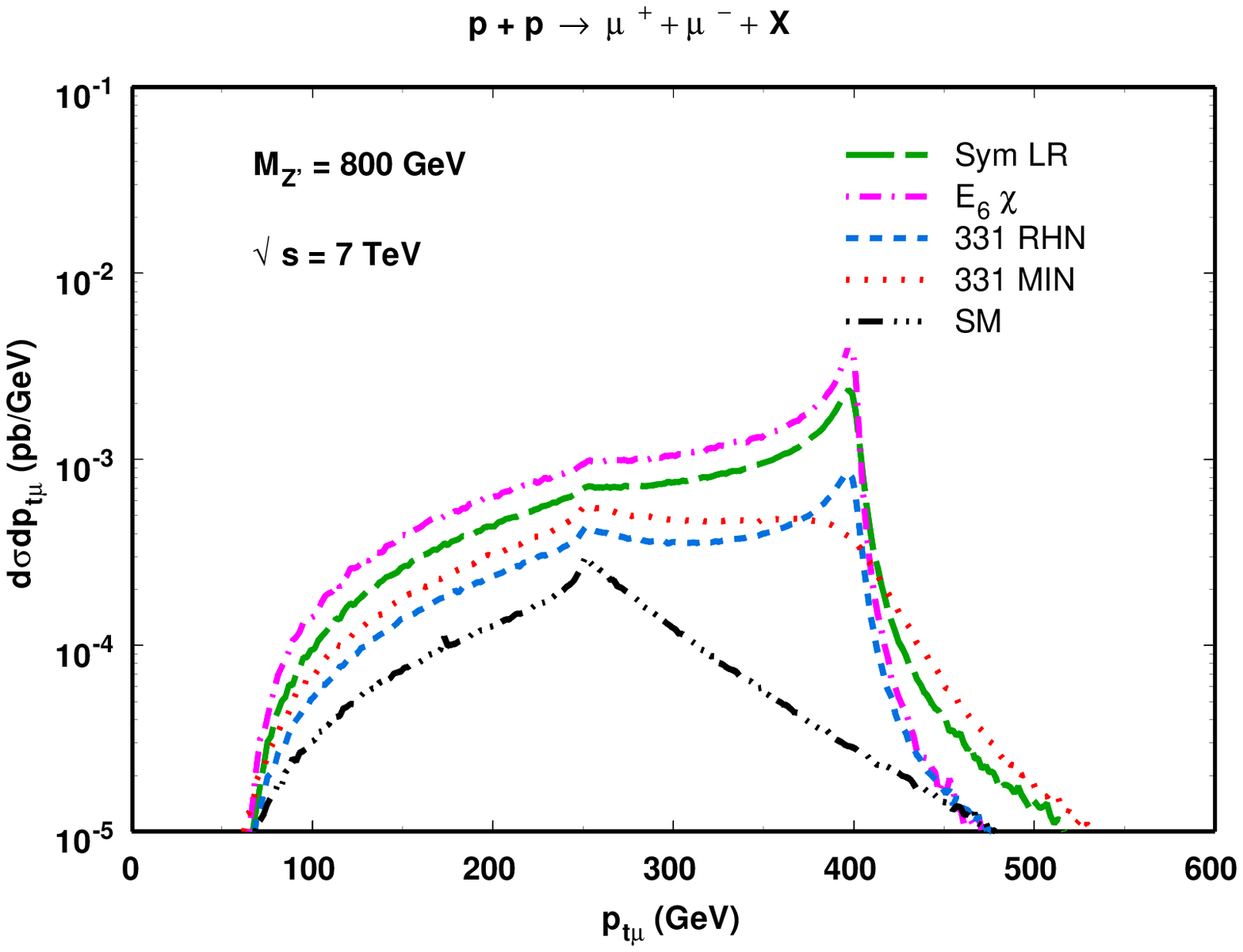}
\hskip -1. cm \includegraphics[height=.3\textheight] {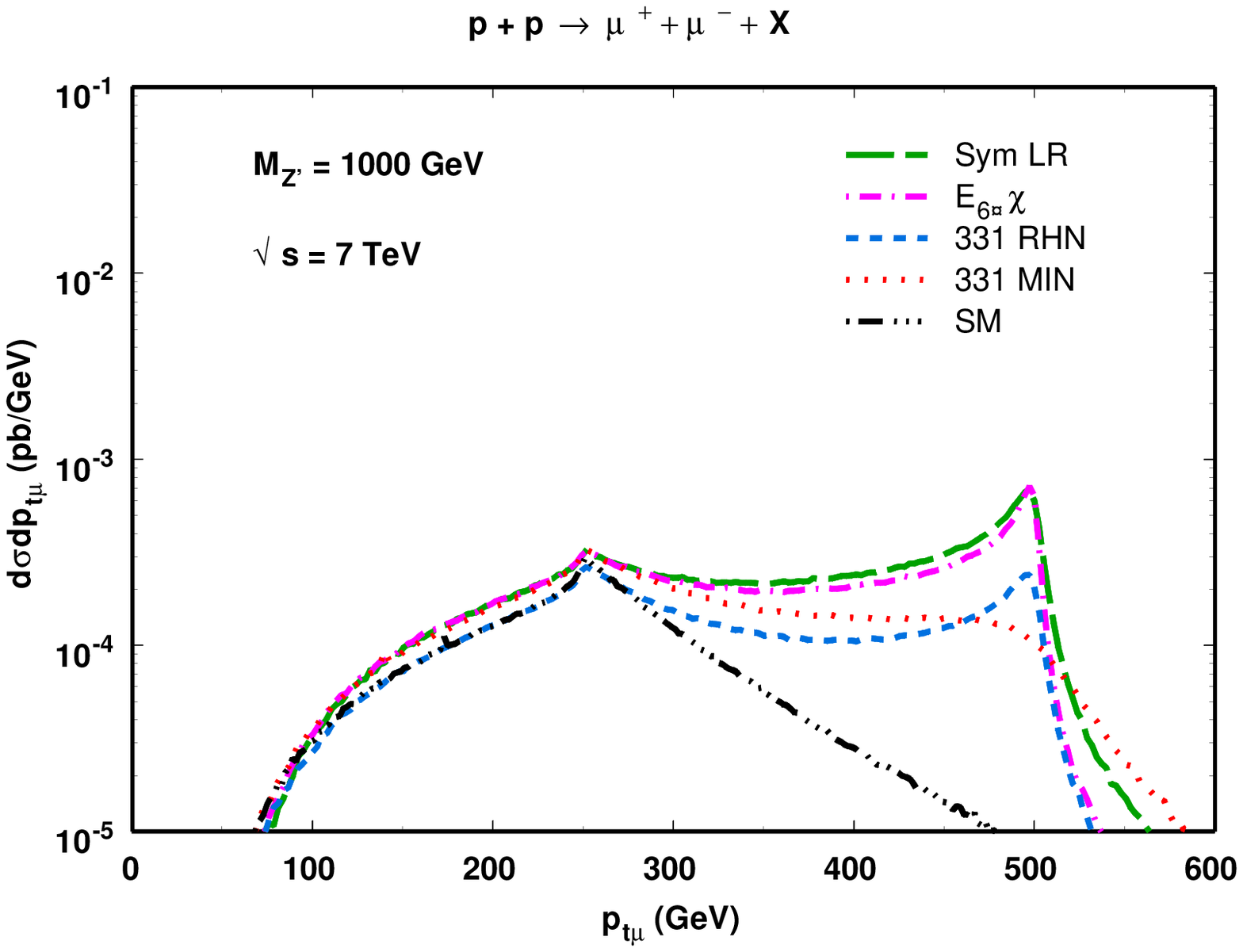}
\hskip -1. cm \includegraphics[height=.3\textheight] {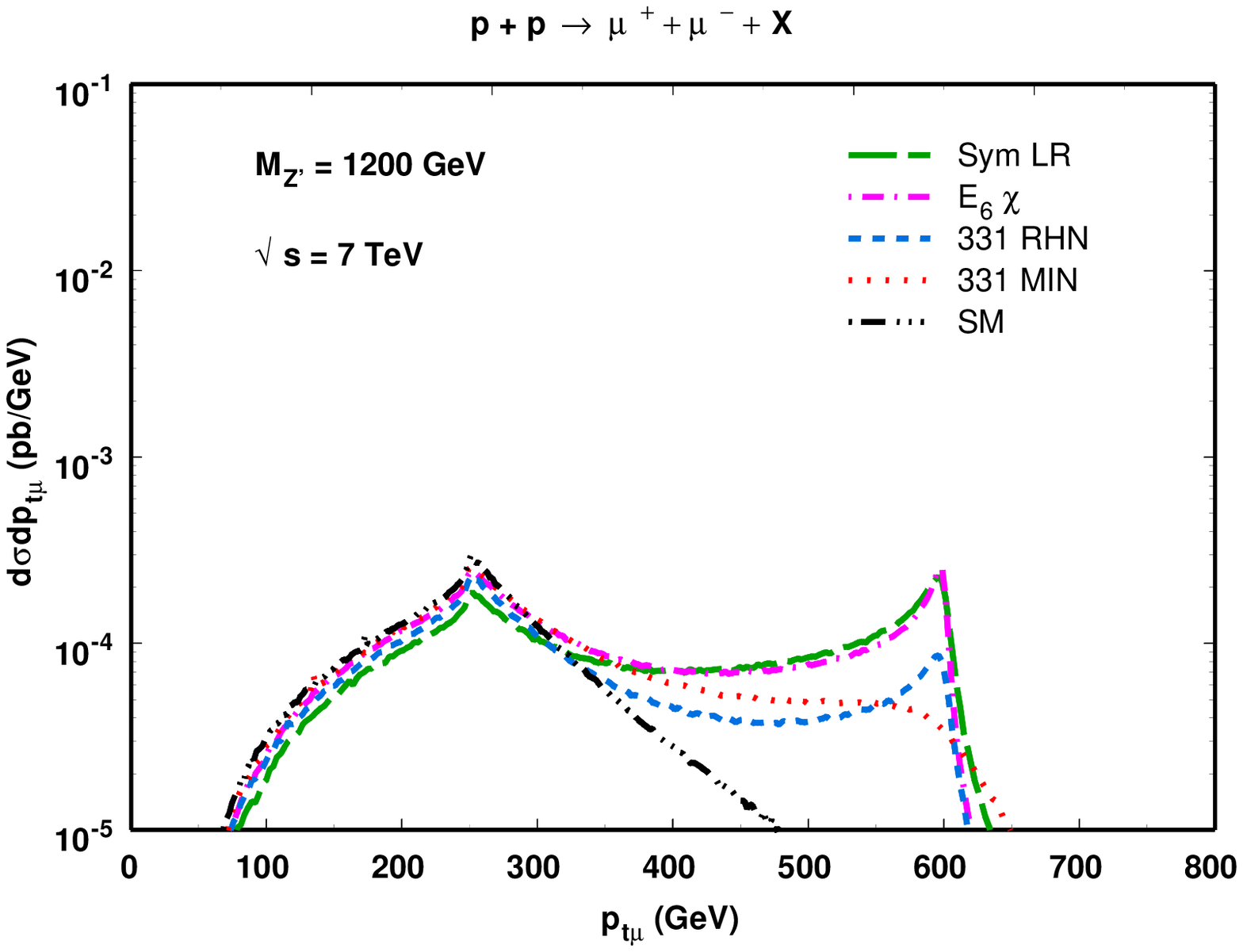}
\caption{\large Lepton transverse momentum distribution for the process $p + p \longrightarrow \mu^+ + \mu^- + X$ ($\sqrt s = 7$ TeV) for some models, considering $M_{Z^{\prime}}=800$ GeV (left), $M_{Z^\prime}=1000 $ GeV (right) and $M_{Z^\prime}=1200 $ GeV (down) and a cut $M_{\mu \, \mu} > 500 $ GeV.}
\end{figure}


\begin{thebibliography}{ABC}
\bibitem{VIP} V.~Pleitez, Phys. Rev. D {\bf 53}, 514 (1996).
\bibitem{PIV} F.~Pisano and V.~Pleitez, Phys. Rev. D {\bf 46}, 410 (1992).
\bibitem{FRA} P.~H.~Frampton, Phys. Rev. Lett. {\bf 69}, 2889 (1992).
\bibitem{RHN} J.~C.~Montero, F.~Pisano and V.~Pleitez, Phys. Rev. D {\bf 47}, 2918 (1993); R.~Foot, H.~N.~Long and T.~A.~Tran, Phys. Rev. D {\bf 50}, R34 (1994); Hoang Ngoc Long, Phys. Rev. D {\bf 53}, 437 (1996); {\it ibid} {\bf 54}, 4691 (1996).
\bibitem{LIT} N.~Arkani-Hamed, A.~G.~Cohen and H.~Georgi, Phys. Lett. B {\bf 513}, 232 (2001) ; N.~Arkani-Hamed, A.~G.~Cohen, E.~Katz and A.~E.~Nelson, JHEP {\bf 0207}, 034 (2002).
\bibitem{LRM} An extensive list of references can be found in R.~N.~Mohapatra and P.~B.~ Pal, "Massive Neutrinos in Physics and Astrophysics", World Scientific, Singapore, 1998.
\bibitem{E6P} P.~Langacker, R.~W.~Robinett and J.~L.~Rosner, Phys. Rev. D {\bf 30}, 1470 (1984). 
\bibitem{E6M} J.~L.~Hewett, T.~G.~Rizzo, Phys. Rep. {\bf 183}, 193 (1989).
\bibitem{RIZ} "Pedagogical Introduction to Extra Dimensions", Thomas G.~Rizzo, SLAC-PUB-10753,  SSI-2004-L013, Sep 2004. hep-ph/0409309.
\bibitem{PDG} C.~Amsler {\it et al.}, Phys. Lett. B {\bf 667}, 1 (2008).
\bibitem{LAN} Jens~Erler, Paul~Langacker, Shoaib~Munir and Eduardo~Rojas~Pe\~na, hep-ph/0906.2435.
\bibitem{RIT} "Z-prime phenomenology and the LHC", Thomas G.~Rizzo, SLAC-PUB-12129, Oct 2006. hep-ph/0610104. 
\bibitem{AGU} F.~del Aguila, M.~Cveti\v{c}, P.~Langacker, Phys. Rev. D {\bf 48}, R969 (1993).
\bibitem{HEP} A.~Pukhov and {\it et al.}, CompHEP - a package for evaluation of Feynman diagrams and integration over multi-particle phase space. Users manual for version 3.3, hep-ph/9908288;
E.~Boos {\it et al.}, [CompHEP Collaboration], CompHEP 4.4: Automatic computations from Lagrangians to events, Nucl. Instrum. Meth. {\bf  A} 534, 250 (2004).
\bibitem{CMS} CMS Technical Design Report, Vol.~II: Physics Performance, J. Phys. G:  Nucl. Part. Phys. {\bf 34}, 995 (2007).
\bibitem{GOD} Stephen Godfrey and Travis A.~W.~Martin, Phys. Rev. Lett. {\bf 101}, 151803 (2008).
\bibitem{ROS} Jonathan L.~Rosner, Phys. Rev. D {\bf 54}, 1078 (1996).
\bibitem{DIT} Michael Dittmar, Phys. Rev. D {\bf 55}, 161 (1997).
\bibitem{DIM} Michael Dittmar, Anne-Sylvie Nicollerat and Abdelhak Djouadi,  Phys. Lett. B {\bf 583}, 111 (2004).
\bibitem{CMS1} Robert Cousins, Jason Mumford, Viatcheslav Valuev, CMS NOTE 2005/022.
\bibitem{VER} J.~Smith, W.~L.~van Neerven, and J.~A.~M.~Vermaseren,  Phys. Rev. Lett.  {\bf 50}, 1738 (1983).
\bibitem{JOH} John R.~Ellis, Mary K.~Gaillard, Georges Girardi, and Paul Sorba, Ann. Rev. Nuc. Part. Sci. 32, 443 (1982).
\bibitem{JEL} John R.~Ellis, CERN-PH-TH/2010-074. hep-ph/1004.0648.
\end{thebibliography}
\end{document}